\documentclass{aastex}
\usepackage{emulateapj5}

\newcommand{\BP}{Ballesteros-Paredes}
\newcommand{\Et}{E_{\rm t}}
\newcommand{\kd}{k_{\rm d}}
\newcommand{\ld}{\lambda_{\rm d}}
\newcommand{\ls}{\lambda_{\rm s}}
\newcommand{\Ms}{M_{\rm s}}
\newcommand{\VS}{V\'azquez-Semadeni}

\shorttitle{}
\shortauthors{}

\begin{document}


\title{A Holistic Scenario of Turbulent Molecular Cloud Evolution and
Control of the Star Formation Efficiency. First Tests}

\author{E.\ V\'azquez-Semadeni\altaffilmark{1}, J.\
Ballesteros-Paredes\altaffilmark{1} and R.\ S.\ Klessen\altaffilmark{2}}

\altaffiltext{1}{Instituto de Astronom\' \i a, UNAM, Campus Morelia, 
Apdo. Postal 3-72, Morelia, Michoac\'an, M\'exico,
e.vazquez, j.ballesteros @astrosmo.unam.mx}

\altaffiltext{2}{Astrophysikalisches Institut Potsdam,
An der Sternwarte 16, 14482 Potsdam, Germany, rklessen@aip.de}

\begin{abstract}
We compile a holistic scenario for molecular cloud (MC) evolution and control
of the star formation efficiency (SFE), and present a first set of numerical
tests of it. A {\it lossy} compressible cascade can generate density
fluctuations and further turbulence at small scales from large-scale motions,
implying that the turbulence in MCs may originate from the compressions that
form them. Below a {\it sonic} scale $\ls$, turbulence cannot induce any
further subfragmentation, nor be a dominant support agent against gravity.
Since progressively smaller density peaks contain progressively smaller
fractions of the mass, we expect the SFE to decrease with decreasing $\ls$, at
least when the cloud is globally supported by turbulence. Our numerical
experiments confirm this prediction. We also find that the collapsed mass
fraction in the simulations always saturates below 100\% efficiency. This may
be due to the decreased mean density of the leftover interclump medium, which
in real clouds (not confined to a box) should then be more easily dispersed,
marking the ``death'' of the cloud.  We identify two different functional
dependences (``modes'') of the SFE on $\ls$, which roughly correspond to
globally supported and unsupported cases. Globally supported runs with most of
the turbulent energy at the largest scales have similar SFEs to those of
unsupported runs, providing numerical evidence of the dual role of turbulence,
whereby turbulence, besides providing support, induces collapse at
smaller scales through its large-scale modes. We 
tentatively suggest that these modes may correspond to the clustered and
isolated modes of star formation, although here they are seen to form part of a
continuum rather than being separate modes. Finally, we compare with previous
proposals that the relevant parameter is the energy injection scale.

\end{abstract}


\section{Introduction} \label{sec:intro}

Two of the main questions about molecular cloud (MC) structure and star
formation concern the origin of molecular cloud turbulence and of the low
efficiency of star formation (SFE). Indeed, the turbulent properties of
molecular clouds are remarkably uniform, even in clouds without stellar
energy sources (see, e.g., McKee 1999 and references therein). And,
defined as the fraction of the mass in a MC that ends up in stars before 
the cloud is dispersed, the SFE is well known to be as 
low as a few percent with respect to the total cloud
mass (see, e.g., Evans 1991 and references therein).

It is known that turbulence can provide additional
support against global gravitational collapse (Chandrasekhar 1951;
Bonazzola et al.\ 1987, 1992; \VS\ \& Gazol 1995). In particular,
numerical studies have suggested that turbulence is able to
prevent the collapse when the energy injection scale is
smaller than the Jeans length at typical core densities
(L\'eorat, Passot \&
Pouquet 1990; Klessen, Heitsch \& Mac Low 2000, hereafter KHM00).  
Simultaneously, supersonic, compressible turbulence is known to produce
``fragmentation'', i.e., the formation of smaller-scale 
density substructures that can themselves possibly undergo local
collapse (Sasao 1973; Hunter \& Fleck 1982; Tohline, Bodenheimer \&
Christodolou 1987; Scalo 1987; 
Elmegreen 1993; Padoan 1995; V\'azquez-Semadeni, Passot \& Pouquet
1996; KHM00; Padoan et al. 2001; Heitsch, Mac~Low \& Klessen
2001). Padoan (1995) and KHM00 have proposed that 
turbulent fragmentation has an important role in determining the
SFE. The latter authors have shown that the efficiency in numerical
simulations (measured as the fraction of mass in collapsed objects)
increases as the energy injection scale is increased. However, this
result was obtained in 
simulations with an idealized random forcing applied at specific scales,
and ubiquitously in time and space. Instead, the
turbulence in real MCs is probably injected over a wide range of scales, 
but at localized sites and times (Scalo 1987; Norman \& Ferrara 1996; 
Avila-Reese \& \VS\ 2001). In this
case, the driving scale is not uniquely defined. Also, if
molecular clouds form through the collisions of larger-scale streams
in the more diffuse ISM (Elmegreen 1993; \BP, \VS\ \& Scalo 1999; \VS,
\BP\ \& Klessen 2002, hereafter Paper I), then they can be part of a
turbulent cascade 
originating from much larger scales, and thus the driving scale can be
much larger than the Jeans length in the molecular 
clouds (Mac Low \& Ossenkopf (2001). This appears inconsistent with the
requirement of a driving scale smaller than the cloud's Jeans length in
order to maintain a low efficiency. 

In Paper I, we have compiled
a holistic scenario in which both the turbulence in molecular clouds and
the SFE are part of the life cycle of MCs: {\it MC turbulence
originates from the very process that forms the clouds}, namely
colliding streams in the diffuse ISM (Elmegreen 1993; \BP\ et al.\
1999), through bending-mode instabilities (Hunter et al.\ 1986; Vishniac
1994; Walder \& Folini 1998; Klein \& Woods 1998). In this sense, MC
turbulence is part of a compressible {\it and lossy} cascade
which may originate at intermediate scales in the Galactic disk
($\gtrsim 1$ kpc) from various
instabilities (Parker 1966; Toomre 1964; Elmegreen 1991),
supernova energy input (e.g., McCray \& Snow 1979; MacLow 2002), and stellar 
winds (Norman \& Silk 1980), and continues towards smaller
scales, as the internal turbulence produces yet smaller-scale turbulent
compressed layers (Kornreich \& Scalo 2000). The compressible 
cascade has to be lossy because shocks are known to be dissipative  
by transferring a fraction of the energy involved directly to the
dissipative scales (Kadomtsev \& Petviashvili 1973). This
compressible-cascade regime, however, 
must end at a ``sonic'' scale, denoted $\ls$, below which the
turbulence internal to the new structures becomes subsonic and therefore
cannot produce any further subfragmentation (Padoan
1995). This does not necessarily imply that the
dissipative (or ``inner'') scale of the turbulence has been reached,
but only that at this point the cascade can turn into an incompressible
one. Thus, in this scenario, 
subsonic ``cores'' are the natural end of the {\it compressible} part
of the cascade in the ISM.  

When the turbulence becomes subsonic, it
ceases to be a stronger source of global support than thermal pressure,
and the Jeans criterion can be applied to determine whether a ``core''
is gravitationally unstable and should proceed to collapse (Padoan
1995), or else 
rebound and merge back into its environment (Elmegreen 1993; Taylor,
Morata \& Williams 1996; \BP\ et al.\ 1999; \VS,
Shadmehri \& \BP\ 2002). This effectively {\it sets an upper limit to
the SFE}, because,  if globally the cloud is supported by its internal
turbulence, only massive enough clumps among those with sizes $l
\lesssim \ls$ are susceptible to local collapse (Padoan 1995). In this
sense, star-forming 
cores constitute the intersection of the set of subsonic regions with
the high-mass, super-Jeans {\it tail} of the mass distribution
(Padoan 1995; \BP\ \& \VS\ 1997; Shadmehri, \VS\ \& \BP\ 2001).
%
%
In particular, Padoan (1995) gave a semi-phenomenological
theory for the efficiency and mass distribution of proto-stellar cores
in turbulent molecular clouds (recently extended by Padoan \& Nordlund
2002). In that work, however, he considered only a power-law
turbulent spectrum, neglecting the necessary existence of inner and
outer scales, and the role of the energy injection scale, which was
proposed to be crucial by L\'eorat et al. (1990) and KHM00. 

%


In this Letter, besides putting together and advancing the above
global scenario, we present (\S\ref{sec:num_exp}) a first set of
numerical tests supporting it, by showing that the scale $\ls$ indeed
plays a crucial role in limiting the SFE. Clearly, this
is only a partial ingredient, as the mass 
distribution in the flow is also expected to be crucial (Padoan
1995). Our experiments also support the notion of the dual role of
turbulence in providing support to large scales while promoting
small-scale collapse, as we discuss in \S\ref{sec:discussion}, where we
also give a re-interpretation of the results of KHM00 in the framework
of our scenario.

\section{Numerical experiments and results} \label{sec:num_exp}

We consider a set of 12 numerical experiments similar to those
reported by KHM00. We refer the reader to that paper for details (see
also Klessen \& Burkert 2000, 2001). Here we just 
mention that they are smoothed particle hydrodynamics (SPH) simulations
of non-magnetic turbulent flows using 205,379 particles, with
self-gravity, and random forcing 
applied only at wavenumbers in a narrow interval $k_{\rm d}-1 \leq |{\bf
k}| \leq k_{\rm d}$, where $k_{\rm d}\equiv L/\lambda_{\rm d}$ is the driving
wavenumber, $L$ is the computational box size and $\lambda_{\rm d}$ is
the driving scale. The simulations are carried out in non-dimensional
units, in which there are 64 thermal Jeans masses (with respect to the
mean density) in the computational box, the time unit is
$t_0=t_{\rm ff}/1.5$, where $t_{\rm ff}$ is the free
fall time, the sound speed is $c=0.1$, and the mean density is
$\rho_0=1/8$. A constant kinetic energy input 
rate is used, with an amplitude chosen to maintain a fixed level of
turbulent kinetic energy $\Et$. For our analysis, the SPH data are
interpolated onto a cubic grid of $128^3$ cells, but note that the
effective resolution of the SPH method in the density peaks (cores) is
higher than that.

The runs are labeled mnemonically as M$x$K$y$, where $x$ denotes the
rms Mach number and $y$ denotes the forcing wavenumber
$\kd$.
The rms Mach number $\Ms$ and the energy injection scale
$\lambda_{\rm d}$ were taken by KHM00 as the controlling parameters in 
their investigation of the amount of mass in collapsed objects $M_*$ as
a function of time for several simulations. The 
numerical scheme allows this mass to be measured through the usage
of ``sink'' particles, which are particles that replace a collapsing gas 
clump once it becomes too dense and small to be properly resolved. A sink
particle inherits the combined masses of the replaced SPH particles, as
well as their linear and angular momenta, and has the ability to
continue accreting further SPH particles. 

In order to investigate the role of the
sonic scale $\ls$, we consider the velocity dispersion-vs.-size
relation, together with the mass-accretion histories of the
simulations. 
The velocity dispersion-size relation we consider is not restricted 
to the densest, or highest-column density projected structures, as is
customary in both oservational (Larson 1981; see Blitz 1993 for a more
recent review) and numerical (\VS, \BP\ \& Rodr\' \i guez 1997;
Ostriker, Stone \& Gammie 2001; \BP\ \& Mac Low 2002) astrophysical
studies. Instead, we consider a subdivision of the whole
computational box in regions of size $l$, and measure the
velocity dispersion in all such regions. The mass
accretion histories are computed as the 
fraction of the total mass in collapsed objects (sink particles) as a
function of time. 

\begin{figure}
\plotone{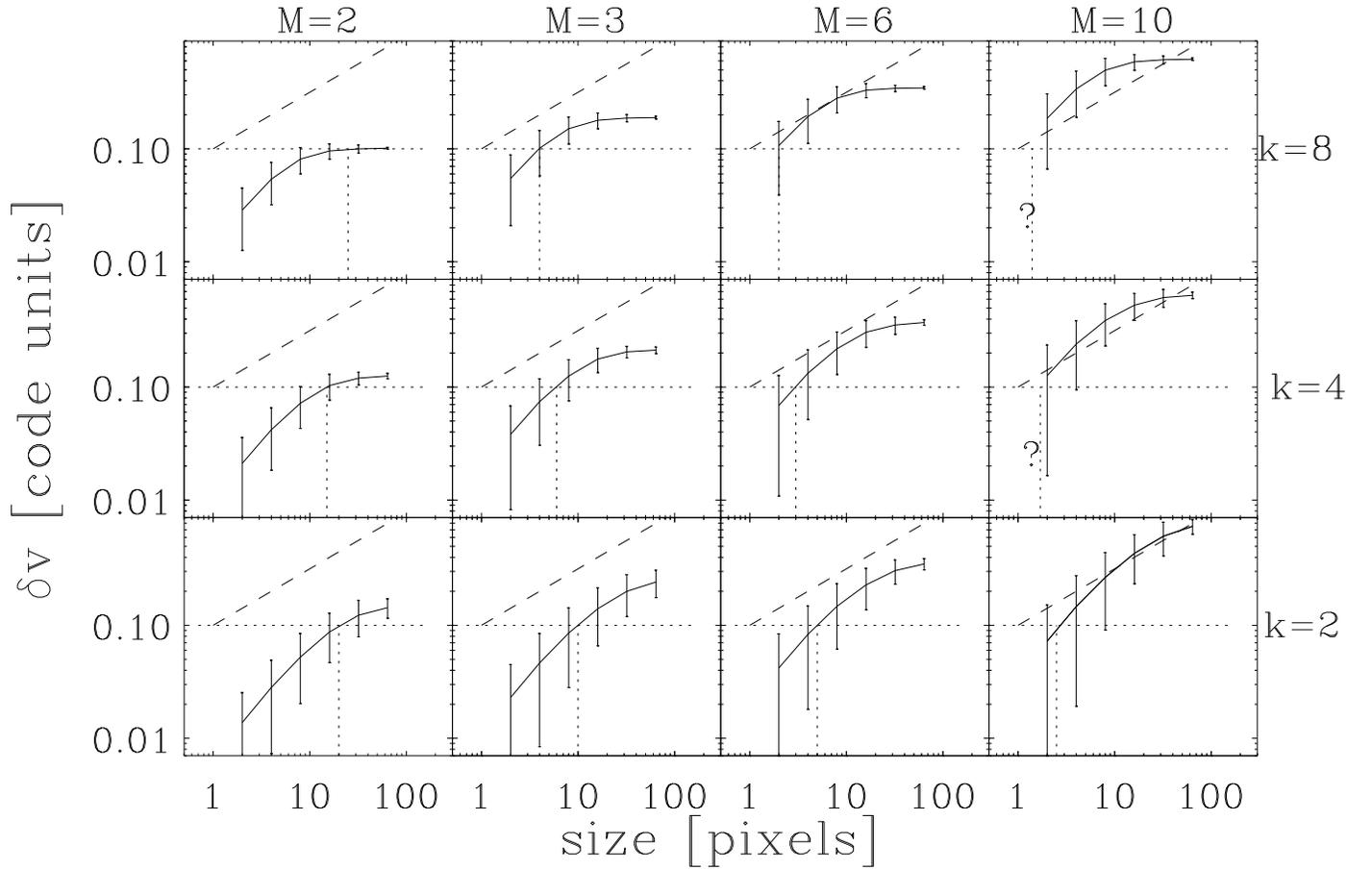}
\caption{Velocity dispersion $\Delta v$ versus size $R$ for all runs
considered in this paper (solid line). The nomenclature of the 
runs indicates their nominal rms Mach number $\Ms$ and the driving
wavenumber $\kd$. 
The dashed line shows an $R^{1/2}$ law, the
horizontal dotted line shows the value of the sound speed, and the
vertical line marks the sonic scale. The error bars indicate the
scatter of the around the mean value of $\Delta v$.}
\label{fig:v_r}
\end{figure}

Figure \ref{fig:v_r} shows the velocity dispersion-size relation for 
all runs. For every run, the solid line shows the average velocity
dispersion for all regions of size $l$; the ``error'' bars indicate the
range of values of the velocity dispersion exhibited by all regions of
size $l$. The dotted straight lines show an $R^{1/2}$
power law, and the dashed horizontal line shows the value of
the sound speed. The sonic scale can be defined as that at which the
solid line crosses the sound speed level, and is depicted by a
vertical dotted line. Because of the
large error (or, more properly, ``scatter'') bars in the velocity
dispersion, the sonic scale is typically defined only to within factors
$\sim 4$. The caveat should be mentioned that the
velocity dispersion-size plots we obtain are clearly not power laws,
with progressively smaller scales being progressively more deficient
in turbulent kinetic energy. This is in part a consequence of the spatially
varying resolution of SPH, with low resolution at low density (in voids)
and high resolution at high density (in cores) (Ossenkopf, Klessen, \& Heitsch
2001), and in part a consequence of a power-law range not being
expected at scales larger than $\ld$ for the small-driving-scale
runs. In any case, the non-power-law 
nature of the spectrum at small scales implies that our measurements
{\it overestimate} $\ls$. We discuss the implications of this in \S
\ref{sec:discussion}. 

\begin{figure}
\plotone{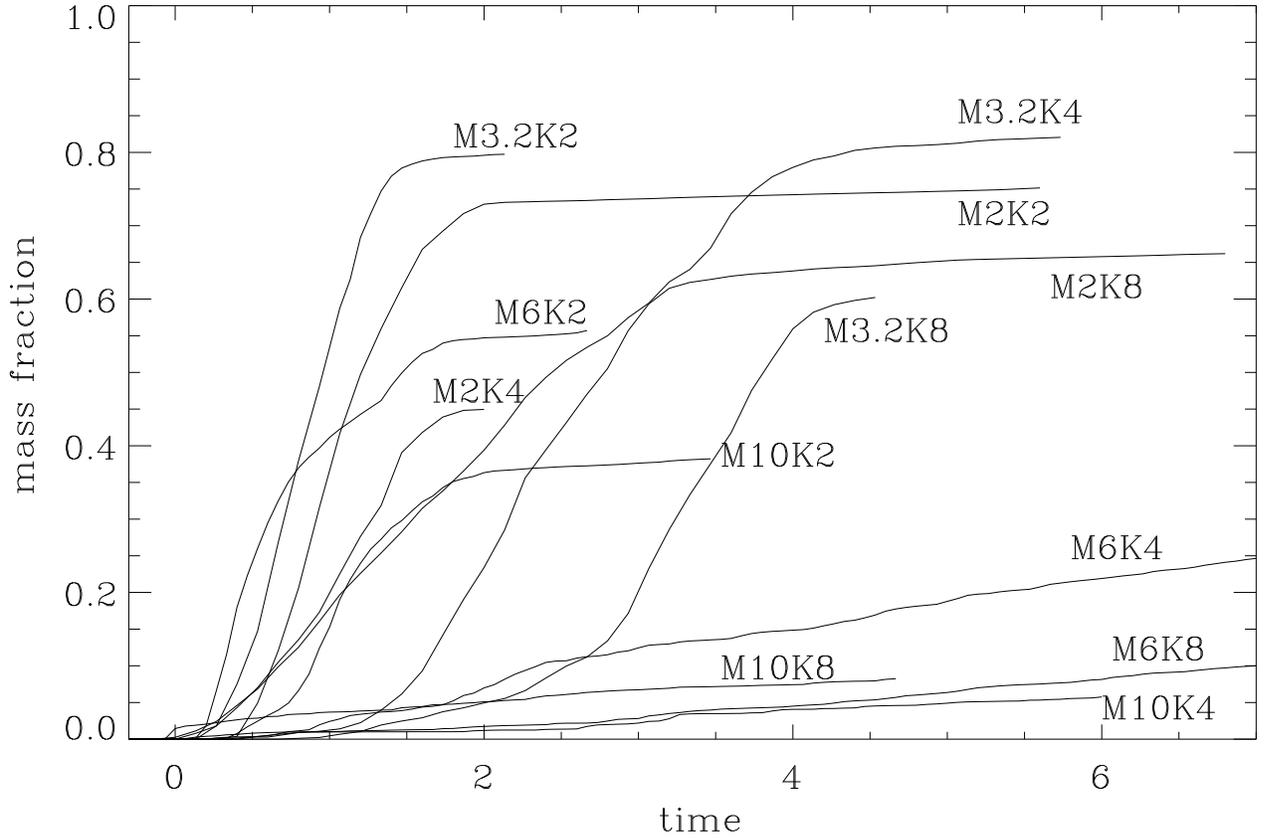}
\caption{Evolution of the collapsed mass fraction for all runs. We
define the star formation efficiency (SFE) as the collapsed mass
fraction at time $t=4$, 6.7 for runs with $\Ms=10$, 6, respectively, and 
as the saturation value for runs with $\Ms$=3.2 and 2.}
\label{fig:accr_hists}
\end{figure}

Figure \ref{fig:accr_hists} shows the accretion histories for all
runs.It is noteworthy that the accreted mass in most runs (except the ones
with the lowest accretion rates) eventually saturates at some
level. This phenomenon is extremely interesting in its own
right; it clearly appears to be a real phenomenon, generally seen
in simulations that allow following the evolution of the system beyond
the formation of individual collapse sites through the use of sink
particles (KHM00; Klessen, Burkert \& Bate 2001; Klessen 2001; Bate,
Bonnell \& Bromm 2002). We speculate
that it occurs because the local collapse events decrease the mean
density of the ``interclump'' medium, making it more difficult to form new 
collapsing sites from it. 
Concerns that this saturation level be a numerical 
artifact dependent of the resolution are dispelled by fig. 6 of KHM00,
which shows no clear trend of the saturation level with resolution, but only 
statistical scatter (different random initial conditions with the same
global parameters show variations of up to a factor of 2 in the final
accreted masses). 

Operationally, defining the SFE is not an unambiguous task, because
not all runs have had time to saturate yet (assuming they all eventually
do). Given that recent studies suggest that clouds have lifetimes of
only a few crossing times (Ballesteros-Paredes et al. 1999; Elmegreen
2000), we choose to define the SFE as the collapsed mass fraction after two
turbulent crossing times ($t_{\rm c}$).\footnote{We thank the referee for
suggseting this criterion.} In the units of the code, these
correspond to $t=10$, 12.5, 6.7 and 4 for runs with rms Mach number
$\Ms=2$, 3.2, 6, and 10, respectively. However, for the runs with
$\Ms=2$ and 3.2, this time is longer than they have been
evolved. Fortunately, the collapsed mass fractions in all these runs
have saturated, and so for them we take the saturation value. Due to the
statistical variation of the collapsed mass fractions mentioned above,
we also put error bars of $\pm 50$\% in the reported values of the SFE.

In order to interpret the results of the simulations, fig.\
\ref{fig:sfe_vs_all} shows the SFE versus $\ld$ ({\it a}), $\Ms$
({\it b}), and $\ls$ ({\it c})
for all simulations. The
linear Pearson correlation coefficients for the logarithmic data
in figs.\ \ref{fig:sfe_vs_all}a, \ref{fig:sfe_vs_all}b and
\ref{fig:sfe_vs_all}c are respectively 0.35, $-0.73$ and 0.75.
showing that indeed the SFE correlates best with the sonic scale
$\ls$. Note that this is in spite of the fact that the relationship
between the SFE and $\ls$ is clearly not a power law, but instead
saturates at large 
$\ls$. A change of variable $\ls \rightarrow u= -0.0038 -0.012
\exp{-2.39 \log \ls}$ gives the fit shown by the solid curve in fig.\
\ref{fig:sfe_vs_all}c, with a much tighter correlation coefficient of
0.86. Nevertheless, the correlation with $\Ms$ 
is also significant. We discuss this in the next section.


\section{Discussion and comparison with previous work} \label{sec:discussion}

\begin{figure}
\plotone{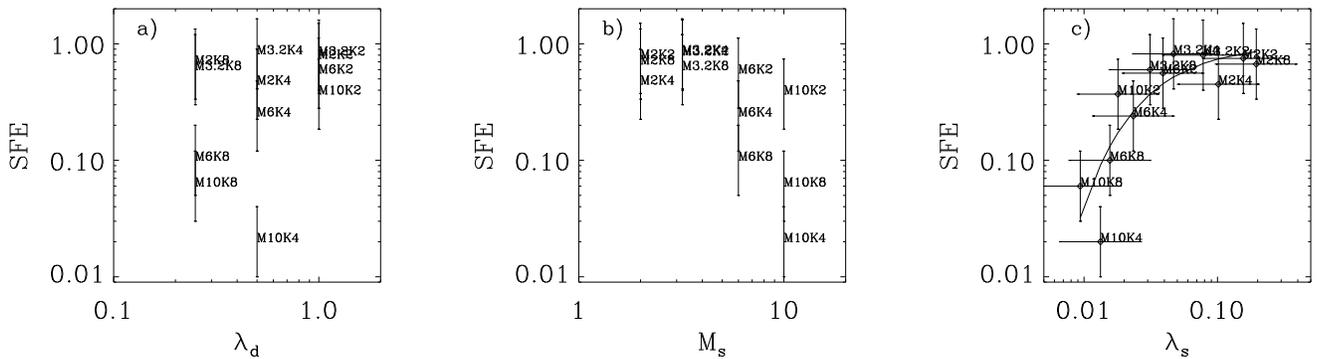}
\caption{Dependence of the SFE on the driving length
$\ld$ ({\it a}), the rms Mach number $\Ms$ ({\it b}) and the sonic
scale $\ls$ in units of the box size ({\it c}).The error bars in the SFE 
indicate the statistical variation among different realizations with
the same global parameters found by KHM00.}
\label{fig:sfe_vs_all}
\end{figure}

Although in the previous section we have fitted a smooth curve to
the data in fig.\  \ref{fig:sfe_vs_all}{\it c}, actually it appears as
if there are two functional dependences (``modes'') of the SFE on
$\ls$ in this plot: for $\ls 
\lesssim 0.03$, the SFE is nearly a power-law on $\ls$,
while it is roughly constant (within a factor of 2) for $\ls \gtrsim
0.03$. All the points in the former regime correspond to cases 
with $\Ms=6$ or 10, which are globally supported by the
turbulence, while most of the points in the latter regime  
correspond to $\Ms=2$ and 3.2, which are globally
unsupported. Nevertheless, the large-scale driven ($\kd = 2$)
high-$\Ms$ cases appear to be at the crossover between the two
regimes. This demonstrates the suggested dual nature of the turbulence
(Sasao 1973; Falgarone, Puget \& P\'erault 1992; \VS\ \&
Passot 1999; KHM00): globally supported runs can be induced to behave
as unsupported ones when the bulk of the turbulent energy is at the
largest scales.

It is tempting to
speculate that these two ``modes'' correspond to the isolated and
clustered observed modes of SF. In this case, our results suggest that
the two would not be physically separated modes, but simply represent
a continuous transition 
from one functional dependence of the SFE on $\ls$ to the other, as the
domain transits from behaving as if globally supported to behaving as
if globally unsupported. We see also that, even in the globally
unsupported case, the efficiency is below unity because of the
saturation described above. In this case, turbulence would only
provide the seeds for the internal fragmentation of the collapsing
molecular cloud region (see, e.g., Klessen et al.\ 1998; Klessen 2001),
the efficiency being more strongly dependent on stellar feedback to
destroy the cloud. Further testing is necessary in order to confirm
the correspondence between the two regimes found here and the
observationally identified isolated and clustered modes.


Concerning the resolution of the numerical
scheme, as mentioned above, $\ls$ is overestimated
at small scales by our simulations. Moreover,
in the absence of numerical diffusion, one would expect the density
peaks to become narrower as well, possibly increasing the number of
them that go into collapse, and thus increasing the SFE. Thus, in
general we expect the correlation seen in fig.\ \ref{fig:sfe_vs_all}c to be
shallower if the diffusivity is decreased. Nevertheless, we do not
expect significant qualitative changes in our results. 

%
%
Our reinterpretation
in terms of $\ls$ eliminates the apparent inconsistency implied
by previous works
that a low SFE would seem to require a small driving scale, yet the
turbulence in molecular clouds appears to be generally driven from
larger scales. Indeed, our results show that a small driving scale alone is
not enough to prevent collapse if it is too 
weak, while driving at large scales can still decrease the efficiency if 
it is strong enough, as pointed out by
KHM00. That is, the controlling parameter is not the driving scale nor
the rms Mach number alone, but instead their combined action expressed
as the energy content at the sonic scale. 
Previous studies did
not consider large-scale-driven cases with a strong enough turbulence
level to decrease the SFE.


A few final remarks: 1) The mechanism
sketched here is expected to operate {\it in addition} to any other mechanism
limiting the SFE based on energy feedback from the massive stars, which
can disrupt the parent cloud. 
2) The occurrence of a saturation of the accreted mass
implies that clouds could not achieve 100\% SFE even if they had
arbitrarily large amounts of time available, provided the turbulence level is
kept constant. This is because {\it local} collapse uses up part of
the mass, rendering the remaining gas more diffuse and easy to
disperse. Thus, {\it partial dispersion may also be part of the natural
course of evolution of molecular clouds}. 3) In this work we have
neglected the effect of the 
magnetic fields. Although their effect certainly must be incorporated in
a full theory, the mechanism discussed here is capable of regulating the
SFE without relying on their presence, depending only on the spatial
redistribution of the physical quantities in a compressible turbulent
flow. This is particularly relevant upon recent suggestions that
molecular clouds may be super-Alfv\'enic (Padoan \& Nordlund 1999) and
that cores are in general supercritical (Nakano 1998; Hartmann,
Ballesteros-Paredes \& Bergin 2001; Bourke et al.\ 2001; Crutcher,
Heiles \& Troland 2002).

\acknowledgements
We thank L.\ Hartmann and the referee, P.\ Padoan, for useful
suggestions. We acknowledge financial support from Conacyt grants 
27752-E to E.V.-S. and I39318-E to J.B.-P., and 
from the Emmy Noether Program of the Deutsche Forschungsgemeinschaft
(KL1358/1) and the Center for Star Formation Studies (NASA-Ames,
UC Berkeley, and UC Santa Cruz), to R.S.K.


\end{document}